\documentclass[aps,prl,twocolumn,showpacs,showkeys,superscriptaddress,floats,floatfix,
preprintnumbers]{revtex4}
\usepackage{amssymb,amsmath}
\usepackage{bm,url,graphicx,subfigure,epsfig}
\usepackage{color}
\usepackage{natbib}
\usepackage{amsmath}
\usepackage{siunitx}
\usepackage{bm,graphicx,subfigure,epsfig}
\usepackage{color}
\usepackage{pxfonts}
\usepackage[T1]{fontenc}
\usepackage{amsfonts}
\usepackage{amssymb}
\usepackage{graphicx}
\usepackage{amsmath}
\usepackage[utf8x]{inputenc}

\providecommand{\U}[1]{\protect\rule{.1in}{.1in}}
\graphicspath{{Pictures/}}
\expandafter\let\csname equation*\endcsname\relax
\expandafter\let\csname endequation*\endcsname\relax
\graphicspath{{Pictures/}}

\newcommand{\beqa}{\begin{eqnarray}}
\newcommand{\eeqa}{\end{eqnarray}}

\begin{document}

\title{Quantum solitons in spin-orbit-coupled Bose-Bose mixtures}
\author{Andrea Tononi}
\affiliation{Dipartimento di Fisica e Astronomia ``Galileo Galilei'', Universit\`a di
Padova, Via Marzolo 8, 35131 Padova, Italy}
\author{Yueming Wang}
\affiliation{Dipartimento di Fisica e Astronomia ``Galileo Galilei'', Universit\`a di
Padova, Via Marzolo 8, 35131 Padova, Italy}
\affiliation{School of Physics and Electronic Engineering, Shanxi University, Taiyuan
030006, China}
\affiliation{Collaborative Innovation Center of Extreme Optics,Shanxi University, 
Taiyuan, Shanxi 030006, China}
\author{Luca Salasnich}
\affiliation{Dipartimento di Fisica e Astronomia ``Galileo Galilei'', Universit\`a di
Padova, Via Marzolo 8, 35131 Padova, Italy}
\affiliation{Istituto Nazionale di Ottica (INO) del Consiglio Nazionale delle Ricerche
(CNR), \\
Via Nello Carrara 1, 50019 Sesto Fiorentino, Italy}

\begin{abstract}
\noindent
Recent experimental and theoretical results show that weakly interacting atomic 
Bose-Bose mixtures with attractive interspecies interaction are stabilized 
by beyond-mean-field effects. Here we consider the peculiar properties of 
these systems in a strictly one-dimensional configuration, taking also into 
account the nontrivial role of spin-orbit and Rabi couplings.
{We show that when the value
of inter- and intraspecies interaction strengths are such that
mean-field contributions to the energy cancel, a self-bound bright
soliton fully governed by quantum fluctuations exists.} 
We derive the phase diagram of the phase transition between 
a single-peak soliton and a multipeak (striped) soliton, produced by the 
interplay between spin-orbit, Rabi couplings and beyond-mean-field effects, 
which also affect the breathing mode frequency of the atomic cloud. 
Finally, we prove that a phase imprinting of the single-peak soliton leads to a 
self-confined propagating solitary wave even in the presence of spin-orbit coupling.
\end{abstract}

\maketitle
\date{\today}

{\it Introduction.}
{Solitons are localized solitary waves propagating with constant shape in a nonlinear medium:}
due to a simple underlining mathematical structure they are ubiquitous in physics, 
with applications to optics \cite{lederer} and hydrodynamics \cite{miles}, from quantum 
field theory \cite{rajaraman} to proteins and DNA \cite{christiansen,yomosa}, 
polymers \cite{heeger}, plasmas \cite{kuznetsov}, and ultracold gases \cite{pitaevski}. 
In the latter field bright solitons emerge as a balance of kinetic energy and 
nonlinear self-interaction in the Gross-Pitaevski equation of the condensate 
\cite{sala} and were first discovered in 2002 \cite{strecker,khaykovich}.

In uniform and weakly interacting Bose-Bose mixtures the crucial role 
of beyond-mean-field quantum fluctuations for the existence of self-bound 
localized states was recently emphasized.
In three-dimensional mixtures with repulsive intracomponent interaction 
and attractive intercomponent one, a mean-field (MF) 
collapsing system is stabilized by the inclusion of 
beyond-mean-field (BMF) effects \cite{petrov2015}, as experimentally 
observed with dipolar systems \cite{kadau,ferrierbarbut,schmitt,chomaz} 
and for isotropic contact interactions \cite{tarruell,semeghini,tarruell2}.
{Contrary to the three-dimensional (3D) case, in a strictly one-dimensional Bose-Bose mixture 
the BMF {\it attractive} energy stabilizes a {\it repulsive} 
MF term \cite{petrov2016}.}

Here we study the one-dimensional quantum bright soliton, namely a
fully quantum self-bound state in which the interparticle 
interactions are tuned to eliminate completely the MF contributions.
{Due to the intrinsic attractive 
nature of the 1D BMF energy an external confining potential is not necessary,
different from the 3D analog of this system \cite{arlt}.
Thus reaching a one-dimensional confinement is truly crucial to 
observe this new self-bound state.}
We investigate the influence of spin-orbit (SO) 
\cite{lin2011,zhai2012,galitski2013,kartashov,malomed} 
and Rabi couplings between the species, deriving a phase diagram 
for the phase transition between a single-peak soliton 
and a striped soliton. Regarding the dynamical properties, 
we calculate the breathing mode frequency of the soliton 
and we find that despite the broken Galilean invariance \cite{zhu}, 
the single-peak soliton propagation is shape invariant.

{\it The model.}
Let us consider a uniform {one-dimensional} Bose-Bose gas made of two species with 
equal mass $m$ and uniform number densities $n_{1}$ and $n_{2}$.
We suppose that the real two-body interaction potential between the atoms 
can be substituted with the same one-dimensional zero-range coupling $g=g_{11}=g_{22}$ 
for intracomponent interactions and with $g_{12}$ for intercomponent ones.
The beyond-mean-field energy density of the mixture reads \cite{petrov2016}
\begin{eqnarray} \label{Eint}
E_{1D}(n_{1},n_{2}) &=&\frac{g}{2}(n_{1}-n_{2})^{2}  \\ &+&  
\frac{\delta g }{4}(n_{1}+n_{2})^{2}
-\frac{2 \sqrt{m}} {3\pi\hbar} g^{3/2}(n_{1}+n_{2})^{3/2}, \nonumber
\end{eqnarray}
where $\hbar$ is the reduced Planck constant and $\delta g = g_{12} + g$.
In particular, we model a weakly interacting mixture near the instability 
point of the mean-field theory, considering the regime of 
$0 \le \delta g \ll g$, with attractive intercomponent interaction 
$g_{12} < 0$ and repulsive intracomponent one $g > 0$. 

Within an effective field theory (EFT), we describe the species with the 
complex scalar bosonic fields $\psi_{1}(x)$ and $\psi_{2}(x)$, thus extending the 
definitions of the uniform particle densities $n_{1}$ and $n_{2}$ 
to the local quantities $n_{1}=|\psi_{1}|^{2}$ and $n_{2}=|\psi_{2}|^{2}$.
In the spirit of density functional theory we introduce the energy functional
\beqa
{\cal E}=\int dx \; &\bigg\{& E_{1D}(|\psi_{1}|^2,|\psi_{2}|^2) + 
\sum_{j=1,2} \bigg[\nonumber{\frac{\hbar^2}{2m}} |\partial_x \psi_j|^2 \\ 
&-& (-1)^j i \gamma \psi_j^* \partial_x\psi_j - \Gamma \psi_j^* \psi_{3-j} \bigg] \bigg\},  
\label{eff1}
\eeqa
which is obtained adding a kinetic energy term to the beyond-mean-field energy 
of Eq. (\ref{Eint}), and including the contributions of an artificial spin-orbit 
coupling with strength $\gamma$ and a Rabi coupling with strength $\Gamma$
between the species.
This low-energy EFT, in our regime of application, is a reliable tool
to determine the static properties of the system \cite{Astra2018}. 
Indeed, the minimization of Eq. (\ref{eff1}) with the chemical 
potential $\mu$ as a Lagrange multiplier fixing the total number of particles 
$N_{1}+N_{2}=2N$ leads to two coupled stationary Gross-Pitaevski
equations (GPE)
\begin{eqnarray}  \nonumber
&\mu \psi_{j} = \bigg[ -{\frac{\hbar ^{2}}{2m}}\partial
_{x}^{2}+\frac{\delta g}{2} 
(|\psi_{1}|^{2}+|\psi_{2}|^{2}) - (-1)^{j}g(|\psi_{1}|^{2}-|\psi_{2}|^{2}) \\
&-\frac{\sqrt{m}} {\pi\hbar} g^{3/2} 
(|\psi_{1}|^{2}+|\psi_{2}|^{2})^{1/2} -(-1)^{j}i\gamma \partial _{x} 
\bigg] \psi _{j} -\Gamma \psi _{3-j}, \; \label{GPE}
\end{eqnarray}
with $j=1,2$.
To study the static properties of the mixture, 
we will focus on the analytical and numerical solution of Eq. (\ref{GPE}) 
{for $N_1=N_2=N$}, considering the case in which the 
beyond-mean-field terms are removed, i.e. $\delta g = 0$.

{\it Quantum bright soliton.}
We now find an analytical solution of the GPE Eq. (\ref{GPE}) within the 
single-field approximation \cite{salasnichmalomed}
\begin{align}
\begin{split} \label{ansatz}
\psi _{1}(x) &= \sqrt{N} \, \phi (x), \\
\psi _{2}(x) &= \sqrt{N} \, \phi ^{\ast }(x).
\end{split}
\end{align}
By substituting it in the coupled GPE, we get the same 
stationary equation for the time-independent complex field $\phi(x)$, namely
\begin{equation}
\mu \ \phi =\left[ -{\frac{\hbar ^{2}}{2m}}\partial _{x}^{2}+i\gamma
\partial _{x} + \delta g N |\phi|^{2} 
-{\frac{\sqrt{2m}}{\pi \hbar }}g^{3/2} N^{1/2} |\phi |\right] \phi
-\Gamma \phi ^{\ast }. \label{1cpGPE}
\end{equation}
This equation can be solved analytically in the absence of spin-orbit 
and Rabi couplings, i.e. if $\gamma =\Gamma =0$ \cite{Astra2018}. 
However, here we investigate the remarkable case where also $\delta g = 0$, 
in which the nonlinearity of Eq. (\ref{1cpGPE}) contains only beyond-mean-field effects
\begin{equation}
\mu \ \phi =\left[ -{\frac{\hbar ^{2}}{2m}}\partial _{x}^{2}-{\frac{\sqrt{2m}%
}{\pi \hbar }}g^{3/2} N^{1/2} |\phi |\right] \phi \;.  \label{1cpGPE2}
\end{equation}%
The 3D analog of this equation, in which quantum fluctuations are not masked 
by mean-field contributions, has been recently investigated \cite{arlt}{, 
although including a confining potential.}
Assuming a real non-negative field $\phi (x)$ {and considering 
that a bright soliton has $\mu <0$}, Eq. (\ref{1cpGPE2}) takes the form 
\begin{equation}
\phi ^{\prime \prime }=-{\frac{\partial W}{\partial \phi }} \quad \quad {\mbox{with}} 
\quad \quad W(\phi )=-\alpha \ \phi ^{2}+\beta \ \phi ^{3},  \label{newton}
\end{equation}%
where each mark $^{\prime}$ represents a derivative with respect to $x$, and
\begin{equation}
\alpha ={\frac{1}{2}}\left( {\frac{2m}{\hbar ^{2}}}\right) |\mu|,  \; \quad %
\quad \beta ={\frac{1}{3}}\left( {\frac{2m}{\hbar ^{2}}}%
\right) {\frac{\sqrt{2m}}{\pi \hbar }}g^{3/2} N^{1/2} \; .
\end{equation}
{The solution of Eq. (\ref{newton}), with vanishing boundary conditions at infinity, is}
\begin{equation} \label{quantumsoliton}
\phi (x)=\phi (0)\; \mbox{sech}^{2}(\sqrt{\alpha /2}\
x) \;,
\end{equation}%
where $\phi (0)={\alpha }/{\beta}$, and the implicit dependence on the 
chemical potential $|\mu|$ is fixed by
imposing the normalization condition $1=\int dx \, |\phi(x)|^2$, obtaining
{$|\mu| = 2^{1/3} m g^{2} N^{2/3}/(3^{2/3} \pi^{4/3} \hbar^2)$}.
We underline that Eq. (\ref{quantumsoliton}) represents a fully 
quantum bright soliton, whose existence is entirely due to 
beyond-mean-field quantum fluctuations.
Moreover, while a GPE equation in 1D with a cubic nonlinearity admits 
a $\mbox{sech}(x)$ solitonic solution \cite{who}, here we consider 
a quadratic nonlinearity and we obtain a solution in the form of 
$\mbox{sech}^{2}(x)$.

{\it Time-dependent variational ansatz.}
We now study the dynamical properties of the quantum bright soliton 
by using a Gaussian time-dependent variational ansatz.
The Bose-Bose mixture dynamics derives from the following effective Lagrangian:
\begin{eqnarray}\label{lagrangian}
{\cal L} = \int dx \; \sum_{j=1,2} \frac{i \hbar}{2} ( \psi_j^* \partial_t \psi_j - 
\psi_j \partial_t \psi_j^*) - {\cal E}
\end{eqnarray}
in which we implicitly introduce the time dependence $t$ in the fields 
$\psi_{1,2}$ and where ${\cal E}$ is given by Eq. (\ref{eff1}).
The low-energy collective excitations of the system can be studied 
analytically with the Gaussian ansatz \cite{salasnich}
\begin{equation} \label{gaussianansatz}
\psi_{1} (x,t)=\psi_{2} (x,t)= \frac{N^{1/2}}{\pi^{1/4} \sigma^{1/2}(t)} \ 
\exp(-\frac{x^2}{2 \sigma^2(t)} + i b(t) x^2) \;,
\end{equation}
where $\sigma(t)$ and $b(t)$ are time-dependent variational parameters.
Substituting the ansatz into Eq. (\ref{lagrangian}) and integrating along 
$x$ one obtains an effective Lagrangian ${\cal L}$ for $\sigma(t)$ and $b(t)$. 
In the absence of SO and Rabi couplings the Euler-Lagrange equation for 
the variational parameter $b$ admits the algebraic solution $b= m \dot{\sigma}/(2 \hbar \sigma)$.
Employing this condition, the Euler-Lagrange equation for the 
Gaussian width $\sigma$ is in a simple harmonic-oscillator form.
In the case of $\delta g = 0$ it can be linearized for small 
perturbations around the equilibrium configuration 
$\sigma_{st} = (3 \pi^{5/6} \hbar^2)/(2^{4/3} m g N^{1/3})$, obtaining 
the oscillatory solution $\sigma(t) = \sigma_{st} + A \cos(\omega_{b} t + \varphi_{0})$,
where $A$ is the oscillation amplitude, $\varphi_{0}$ is an integration 
constant and $\omega_{b}$ is the breathing mode frequency of the quantum soliton,
which is given by
\begin{equation} \label{omegab}
{\omega_{b} = \frac{2^{13/6}}{3^{3/2} \pi^{5/3}} 
\frac{m}{\hbar^3}  N^{2/3} g^2.}
\end{equation}
In the numerical part we will compare the quantum soliton oscillation frequency 
with the analytical result for $\omega_{b}$. Moreover, we will see that an 
oscillatory behavior characterizes also the low-energy excitations 
of the quantum bright soliton in the presence of nonzero SO and Rabi couplings.
Even though the ground-state solution of Eq. (\ref{quantumsoliton}) is not in a Gaussian 
form, we will show that our ansatz of Eq. (\ref{gaussianansatz}) 
gives a better result than an analogous procedure with 
$\psi_{1,2} \propto \mbox{sech}^{2}(\sqrt{\alpha/2} \, x) \, \mbox{e}^{i b x^2}$,
which leads to {
$\omega_{b}^{'}= c \, \omega_{b}$, with $c=(3^{11/6}\pi^{1/3})/(2^{5/6}5^{1/2}(\pi^2-6)^{1/2}) \approx 1.4$}. 

{\it Numerical results: static properties.}
The ground state of the system is obtained through a two-component predictor-corrector 
Crank-Nicolson algorithm, {which solves Eqs. (\ref{GPE}) with the formal 
substitution $\mu \to -\hbar \partial_{\tau}$, where $\tau$ is the imaginary time.}
The evolution of an initial discretized spinor state $(\psi_1 \;  \psi_2)$ 
is performed and the wavefunctions are renormalized at each time step \cite{chiofalo}.
We stress that, in presence of SO and Rabi couplings, the imaginary time 
dynamics of the algorithm is highly dependent on the phase of the initial conditions 
and can converge to local minima of the energy instead of the absolute one \cite{recati}. 
{Therefore, to reach the ground state,} we take as initial condition for both 
components a Gaussian centered in $x=0$ and {width} $\sigma=2$. 

{Following a standard approach \cite{salasnichparola,salasnichmalomed}, we rescale the 
lengths in units of the characteristic length $l_{\bot}=\sqrt{\hbar/(m\omega_{\bot})}$ 
of the transverse harmonic confinement with frequency $\omega_{\bot}$.
The system is strictly one-dimensional only if the transverse width of the bosonic 
sample is equal to $l_{\bot}$ \cite{salasnich}.
Consistently, here we rescale time in units of $\omega_{\bot}^{-1}$, while $g$, 
$\delta g$, $\gamma$ are in units of $\hbar^2/(m l_{\bot})$, and $\Gamma$ is in 
units of $\hbar \omega_{\bot}$.
We point out that, in a macroscopic system with $N \gg 1$, the mean-field contribution of the 
intraspecies interaction in Eqs. \ref{GPE} is negligible and the relevant interaction 
term is the beyond-mean-field one, which scales with $g N^{1/3}$.
}
\begin{figure}[hbtp]
\includegraphics[scale=0.54]{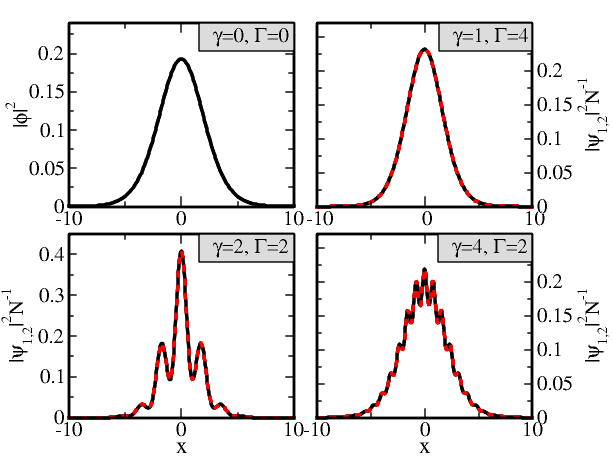}
\caption{Density distribution of the quantum bright soliton ($\delta g =0$)  for different 
values of spin-orbit $\gamma$ and Rabi $\Gamma$ couplings, 
obtained for a fixed intraspecies interaction $g N^{1/3}=1$. 
In the top-left panel we show the square modulus $|\phi|^2$ of our analytical solution 
Eq. (\ref{quantumsoliton}), exactly coincident with the numerical solution of the coupled 
Eqs. (\ref{GPE}) for $\delta g = 0$, $\gamma = 0$, and $\Gamma=0$.
In the other panels we report the normalized components $|\psi_{1}|^2 N^{-1}$ 
(black line) and $|\psi_{2}|^2 N^{-1}$ (red dashed line) with nonzero $\gamma$ 
and $\Gamma$, which turn out to be coincident. 
Here the axial coordinate $x$ is rescaled in units of the transverse 
harmonic-oscillator length $l_{\bot}=\sqrt{\hbar/(m\omega_{\bot})}$, 
with $\omega_{\bot}$ the transverse frequency of the confining potential, 
while $g$, $\delta g$, and $\gamma$ are in units of 
$\hbar^2/(m l_{\bot})$, and $\Gamma$ is in units of $\hbar \omega_{\bot}$.
}
\label{fig:1}
\end{figure}

{The top-left panel of Fig. \ref{fig:1} shows the density profile from
numerical simulations for $\gamma=\Gamma=0$. The profile is
indistinguishable from the analytical prediction in Eq. (\ref{quantumsoliton}).
We have verified that, for $\Gamma=0$, the square modulus of the wave functions does 
not depend on $\gamma$, as previously shown in Ref. \cite{salasnichmalomed}.
This is due to the fact that the spin-orbit coupling can be reabsorbed 
in a phase shift of the fields.}
The other panels of Fig. \ref{fig:1} show the interplay between the 
spin-orbit $\gamma$ and Rabi $\Gamma$ couplings. 
The qualitative effect of SO is to split the bright soliton into many peaks. 
In particular, tuning $\gamma$ from values lower than $\Gamma$ to greater 
ones a larger number of peaks is obtained, but with a finer spatial 
distribution and a smaller density displacement.
Figure \ref{fig:1} also shows that the two components 
have the same ground state distribution, underlining the effectiveness of 
a single-field approximation in the study of attractive Bose-Bose mixtures.

\begin{figure}[hbtp]
\centering
\includegraphics[scale=0.5]{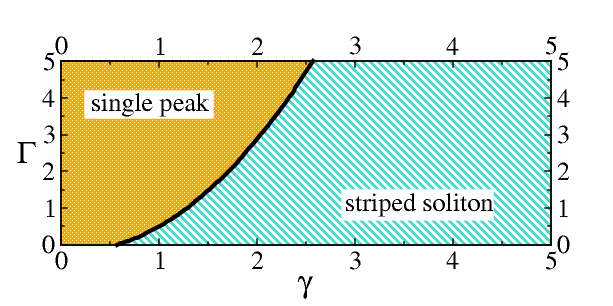}
\caption{Phase diagram of the phase transition from a single peak quantum 
bright soliton to a striped (multipeak) soliton, obtained for a 
fixed intraspecies interaction strength $gN^{1/3}=1$.
The couplings $g N^{1/3}$, $\gamma$, and $\Gamma$ are rescaled as in Fig. \ref{fig:1}.}
\label{fig:2}
\end{figure}
In Fig. \ref{fig:2} we show the phase diagram of the quantum bright 
soliton for the intraspecies interaction coupling $g N^{1/3}=1$. 
The top-left part of the diagram is where the quantum bright soliton has a 
single-peak shape, while in the bottom-right one gets a striped bright soliton, as 
can be seen in comparison with Fig. \ref{fig:1}. The transition black line is given 
by the equation $\Gamma=-0.17-0.19 \gamma + 0.85 \gamma^2$, obtained with a 
polynomial fit of the transition points in the $(\Gamma,\gamma)$ plane:
{this curve characterizes a quantum phase transition fully driven by spin-orbit and 
Rabi couplings.}

{\it Numerical results: dynamical properties.}
The dynamics of the quantum bright soliton is investigated through the 
solution of the following coupled Gross-Pitaevski equations
\begin{eqnarray}  \nonumber
&i \hbar \partial_{t} \psi_{j} = \bigg[ -{\frac{\hbar ^{2}}{2m}}\partial
_{x}^{2}+\frac{\delta g}{2} 
(|\psi_{1}|^{2}+|\psi_{2}|^{2}) - (-1)^{j}g(|\psi_{1}|^{2}-|\psi_{2}|^{2}) \\
&-\frac{\sqrt{m}} {\pi\hbar} g^{3/2} 
(|\psi_{1}|^{2}+|\psi_{2}|^{2})^{1/2} -(-1)^{j}i\gamma \partial _{x} 
\bigg] \psi _{j} -\Gamma \psi _{3-j}, \; \label{tdGPE}
\end{eqnarray}
which are the Euler-Lagrange equations of the Lagrangian 
(\ref{lagrangian}). In particular, we study the breathing 
mode frequency $\omega_{b}$ after an excitation 
of the quantum bright soliton \cite{salasnich}.

\begin{figure}[hbtp]
\includegraphics[scale=0.54]{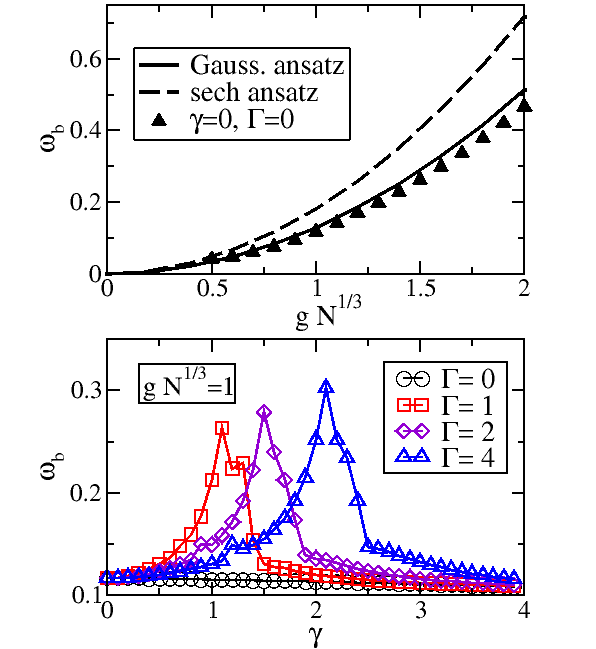}
\caption{Top panel: breathing mode frequency $\omega_{b}$ as a function of $g N^{1/3}$, 
for $\gamma=0$ and $\Gamma=0$.
The symbols are obtained solving numerically Eq. (\ref{tdGPE}),
the black solid line represents our Gaussian ansatz Eq. (\ref{omegab}), 
while the green dashed line is obtained with a 
$\mbox{sech}^{2}(x)$ ansatz (see text). 
{Bottom panel: $\omega_{b}$ as a 
function of $\gamma$, for many values of $\Gamma$, and $g N^{1/3}=1$. 
Notice that $\omega_{b}$ increases at the transition between a 
single-peak and a multipeak soliton. The solid lines are a guide to the eye.}
Here we rescale $\omega_b$ in units of the transverse frequency 
$\omega_{\bot}$, while $t$ is in units of $\omega_{\bot}^{-1}$ 
and all the remaining couplings are rescaled as in Fig. \ref{fig:1}.
}
\label{fig:3}
\end{figure}

In the top panel of Fig. \ref{fig:3} we report $\omega_{b}$ as a function of the intraspecies 
interaction strength $gN^{1/3}$ for fixed values of $\gamma$ and $\Gamma$.
The numerical simulation for $\gamma=0$ and $\Gamma=0$ shows 
a $g^{2}N^{2/3}$ dependence of the breathing mode frequency, 
and is reproduced by our Gaussian ansatz of Eq. (\ref{omegab}) within a 
$9\%$ relative error for $g N^{1/3} \in [0.5,2]$. 
{As previously shown, an analogous calculation of $\omega_{b}$} with a variational 
$\mbox{sech}^2(x)$ ansatz gives the same proportionality to $g^{2}N^{2/3}$, 
but a different coefficient.
{We stress that,} although the soliton density is not a Gaussian, the Gaussian 
ansatz captures the correct oscillatory behavior of the quantum bright soliton.
{In the bottom panel of Fig. \ref{fig:3} we show 
how the breathing mode frequency changes for tuning $\gamma$ 
with $\Gamma$ and $g N^{1/3}=1$ fixed. We find an increase of 
$\omega_{b}$ at the phase transition between a single-peak and a striped soliton: 
this dynamical behavior is a simple experimental test to observe this quantum 
phase transition.}
Notice that we only report the results for one component{, 
since the two species oscillate in time with the same frequency 
and in opposition of phase, such as the center of mass remains always at $x=0$.}

\begin{figure}[hbtp]
\centering
\includegraphics[scale=0.4]{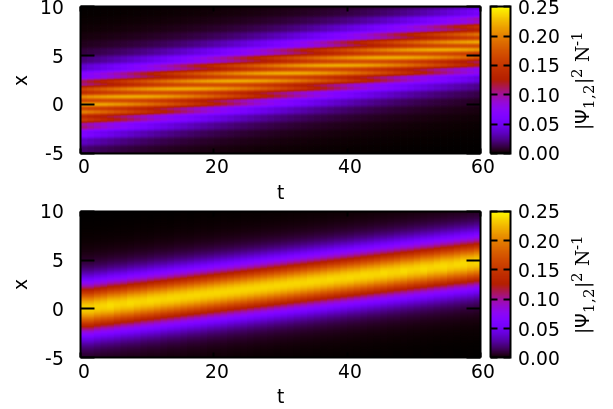}
\caption{Time evolution of the quantum bright soliton after a phase imprinting in the 
form of $\exp(i k x)$, with wavevector $k=2 \pi /60$. 
The unperturbed initial conditions are the striped soliton with 
$g N^{1/3}=1$, $\gamma=4$, and $\Gamma=2$ (top panel) and the single-peak soliton for 
$g N^{1/3}=1$, $\gamma=1$, and $\Gamma=4$ (bottom panel): 
notice how the Galilean invariance is violated only for the striped soliton.
The physical quantities are rescaled as in Fig. \ref{fig:1} and Fig. \ref{fig:3}.
}
\label{fig:4}
\end{figure}
Finally, we analyze the effect of a phase imprinting of the quantum bright 
soliton, which consists in a sudden quench of the phase of the mixture \cite{denschlag}.
Given the stationary ground-state solution $(\psi_1,\psi_2)$, {with Eq. 
(\ref{tdGPE})} we perform the time evolution of the shifted state 
$(\exp(i k x) \, \psi_1,\exp(i k x) \, \psi_2)$, where $k$ is a constant wavevector.  
{
With this phase imprinting the soliton moves with the velocity $v=\hbar k/m$.
To avoid the excitation of transverse modes, which will make the system no longer  
one dimensional, we choose a kick with an energy $\hbar^2 k^2/(2m)$ much smaller 
than the energy of the transverse confinement $\hbar \omega_{\bot}$.}
Our striped soliton is not shape-invariant: as can be seen in the top 
panel of Fig. \ref{fig:4}, during the time evolution in which the 
fluid drifts along $x$, the smaller density peaks do not move.
This is not surprising, because in the presence of SO coupling, the equations 
are not Galilei invariant \cite{zhu}. 
However, we find that the single-peak soliton (bottom panel) propagates 
without changing its shape even with a nonzero SO coupling. 
This is due to the fact that the initial wavefunction is real.

{\it Conclusions.}
We have obtained, {choosing the interaction strength parameters in a way that 
the mean-field terms in the Gross-Pitaevski equation add to zero},
an analytical expression of the quantum bright soliton, namely a self-bound 
structure which can be experimentally observed only in a strictly one-dimensional 
Bose-Bose mixture. 
We have analyzed the phase diagram of the phase transition driven by 
the interplay of spin-orbit $\gamma$, and Rabi $\Gamma$ couplings, which produce
either a single-peak soliton for $\gamma \ll \Gamma$ or a striped soliton for 
$\gamma > \Gamma$.
{Up to now, the only bosonic system with spin-orbit coupling 
realized in the experiments is $^{87}\text{Rb}$. 
Unfortunately, for this species it is truly difficult to tune the intracomponent 
scattering lengths \cite{marte}: this is instead possible with $^{39}$K atoms 
\cite{roy,tanzi}, as recently demonstrated in 3D experiments \cite{tarruell,tarruell2}. 
We suggest this atomic sample as a possible tool to realize in the near future 
quantum bright solitons with spin-orbit coupling, overcoming the difficulties 
expected from the heating of the cloud by the Raman beams \cite{pwang,privcomm}.}

{Let us consider $N \approx 2 \times 10^5$ atoms in different hyperfine levels of $^{39}$K, 
confined in a 1D configuration with the very strong harmonic confinement 
$\omega_{\bot}=4\pi  \times 10^3 \, \text{s}^{-1}$.}
{For $\gamma/(\hbar^2/m)= 6 \times 10^6 \, \text{m}^{-1}$, 
the three-dimensional scattering lengths $a_{11}=a_{22}= 40 \, a_0$, and 
$a_{12}=- 50 \, a_0$, where $a_0$ is the Bohr radius, the transition 
from a multipeak quantum soliton to a single peak one can be 
observed by tuning $\Gamma/(2 \pi \hbar)$ from $1 \times 10^3 \, \text{s}^{-1}$ 
to $4 \times 10^3 \, \text{s}^{-1}$ \cite{privcomm}.
We stress that our simulations show that the transition between a single 
and a multipeak is qualitatively unchanged for $a_{11}=a_{22} \approx -a_{12}$.
Moreover, under these conditions the system 
is very far from the confinement induced resonance \cite{olshanii}, 
since $l_{\bot} \approx 4700 \, a_{0}$ is much larger than all of the suggested 
values of the s-wave scattering lengths.
The present work paves the way to the study of other fully quantum nonlinear excitations, 
like dark solitons, quantized vortices, and shock waves.}

\begin{acknowledgments}
{We thank C. R. Cabrera, A. Simoni, and L. Tarruell for useful discussions.}
The author Y. Wang acknowledges partial support by China Scholarship 
Council, Shanxi 1331KSC and 111 Project (No. D18001). 
\end{acknowledgments}

\end{document}